# Growth of high-quality $CrI_3$ single crystals and engineering of its magnetic properties via V and Mn doping

Shuang Pan, Yuqing Bai, Jiaxuan Tang, Peihao Wang, Yurong You, Guizhou Xu[*], Feng Xu[*]

MIIT Key Laboratory of Advanced Metallic and Intermetallic Materials Technology, School of Materials Science and Engineering, Nanjing University of Science and Technology, Nanjing 210094, China

**Abstract**

$CrI_3$, as a soft van der Waals layered magnetic material, has been widely concerned and explored for its magnetic complexity and tunability. In this work, high quality and large size thin $CrI_3$, V and Mn doped single crystals were prepared by chemical vapor transfer method. A remarkable irreversible Barkhausen effect was observed in $CrI_3$ and $CrMn_{0.06}I_3$, which can be attributed to the low dislocation density that facilitates movement of the domain walls. In addition, the introduction of the doping element Mn allows higher saturation magnetization intensity. $Cr_{0.5}V_{0.5}I_3$ exhibits substantially increased coercivity force and larger magnetocrystalline anisotropy compared to $CrI_3$, while kept similar Curie temperature and good environmental stability. The first principles calculations suggest direct and narrowed band gaps in $Cr_{0.5}V_{0.5}I_3$ and $VI_3$ comparing to $CrI_3$. The smaller band gaps and good hard magnetic property make $Cr_{0.5}V_{0.5}I_3$ an alternative choice to future research of spintronic devices.

---

[*] Corresponding Author: E-mail: gzxu@njust.edu.cn
[*] Corresponding Author: E-mail: xufeng@njust.edu.cn

# 1. Introduction

Two-dimensional (2D) layered ferromagnetic (FM) materials have been the subject of extensive research due to their highly tunable physical properties that show great potential in fundamental physics and next generation spintronics [1-4]. Typical examples of 2D layered FM materials are $Cr_2Ge_2Te_6$ and $CrI_3$ [5-7], where the Curie temperatures ($T_c$) of bulk and monolayer $CrI_3$ are around 61 K and 45 K, respectively [8, 9]. As a layered material, $CrI_3$ exhibits 3D-Ising magnetism [10, 11], low dissociation energy [12], high in-plane stiffness and easy exfoliation [9, 13], maintaining good and stable ferromagnetism even after exfoliation into monolayers [3]. It is interesting to note that the magnetic properties of $CrI_3$ depend on the number of layers: with mono-layer and bulk being ferromagnetic (FM), while double-layers exhibiting antiferromagnetic (AFM) behavior and multilayers being the coexistence of FM and AFM state [14-16]. The unique magnetic transition provides and increases the opportunity and possibility to explore and design low-dimensional intrinsic ferromagnetic semiconductors.

In recent years, in addition to the studies on layer control of $CrI_3$, great efforts have been made to modify its magnetic behavior, such as hydrostatic pressure [17, 18] and electrostatic doping [19, 20]. The application of hydrostatic pressure can alter the interlayer magnetism by influencing the form and strength of the interlayer magnetic coupling. The electrostatic doping is achieved by applying voltage on $CrI_3$-based heterostructures. Both methods require elaborate fabrication of microdevices, which is complex and not easy to operate. There have also been theoretical predictions based on first principles calculations and Monte Carlo simulations that the electron or hole doping can enhance the FM stability of $CrI_3$ as well as its Curie temperature [21]. However, there is still rare experimental investigation about the introduction of foreign ions in $CrI_3$ to tune its magnetic behavior. Also in most of the current studies of $CrI_3$, the preparation method still follows the one proposed in ref. [22], which is modified here to increase the quality of the single crystals while greatly reducing the preparation time.

In this work, a high-quality and large-size $CrI_3$ single crystal was prepared in a relative short time with the method of chemical vapor transport (CVT). By comprehensive study of its structure and physical properties, a significant Barkhausen effect was observed during a period of increasing magnetic field. Mn and V ions are successfully introduced to the matrix $CrI_3$ phase by similar method. It is found that V doped sample ($Cr_{0.5}V_{0.5}I_3$) reveals larger magnetocrystalline anisotropy and higher coercivity force ($H_c$) compared to CrI$_3$, while Mn doping can possibly affect the structure of CrI$_3$ and enhance the average magnetic moment of Cr ion, leading to the higher saturation magnetization. Our results could possibly provide better choices when constructing spintronic devices based on CrI$_3$.

## 2. Experimental section

### 2.1 Sample preparation

Thin CrI$_3$ were synthesized by chemical vapor transfer reaction, using chromium powder and anhydrous iodine beads as raw materials in a molar ratio of 1:3, sealed in a vacuumed quartz tube. The length, inner diameter and wall thickness of the quartz tubes were 19.5 cm, 16 mm and 2 mm, respectively. The quartz tube was placed in a dual-temperature zone tube furnace with one side was set at 923 K, where a total mass of 0.5 g of raw material was put, and the other side where the crystal would grow was kept at a temperature of 623 K. This temperature gradient was maintained for 3 days and then the tube was cooled to room temperature. Finally, a large-size of CrI$_3$ single crystal (about 9 mm in dimension) was formed, as seen in Fig. 1, with the thickness in the range of 1-5 μm.

Thin CrI$_3$ as a layered van der Waals (vdW) material exhibits a monoclinic phase with space group *C2/m* at room temperature and a rhombohedral phase with space group $R\bar{3}$ at low temperature, and the corresponding phase transition temperature is reported to be about 210-220 K [14, 22, 23]. In the monoclinic phase, each layer of Cr$^{3+}$ is coordinated by six nonmagnetic I$^-$ to form an octahedron, which further shares edges to build a honeycomb network. In the rhombohedral phase, the structure consists of three layers of I-Cr-I stacked along the *c\**-axis, and the three adjacent layers are bonded by weak vdW forces to form a honeycomb lattice [24-26].

The V and Mn doped samples were both obtained by similar method, only the atomic ratio and low temperature end changes to Cr:V/Mn:I of 0.5:0.5:4 and 723 K, respectively. After scanning their composition by EDS, it is noted that despite the obtained V doped CrI$_3$ has the designed proportion, the atomic ratio of Mn is only 0.06 (generally below 0.1). By repeated measurements, the existence of Mn atom is confirmed, and the ratio varies between 0.04-0.08. The low solubility of Mn in CrI$_3$ may be originated from that there is no isostructural phase of MnI$_3$ to CrI$_3$, hence the extra Mn is assumed to occupy the interstitial sites, like the vdW gaps. But for V doping, as VI$_3$ has the same structure to CrI$_3$, they are supposed to substitute each other more easily. The size of the doped samples was reduced to a maximum lateral size of about 4-8 mm or even smaller, but the sample thickness was almost unchanged. For VI$_3$, the growing temperatures in the high and low temperature regions were changed to 673 K and 493 K, respectively. The maximum dimension of VI$_3$ single crystal can reach to 1 cm, but it has lower stability and easier to dissolve in air.

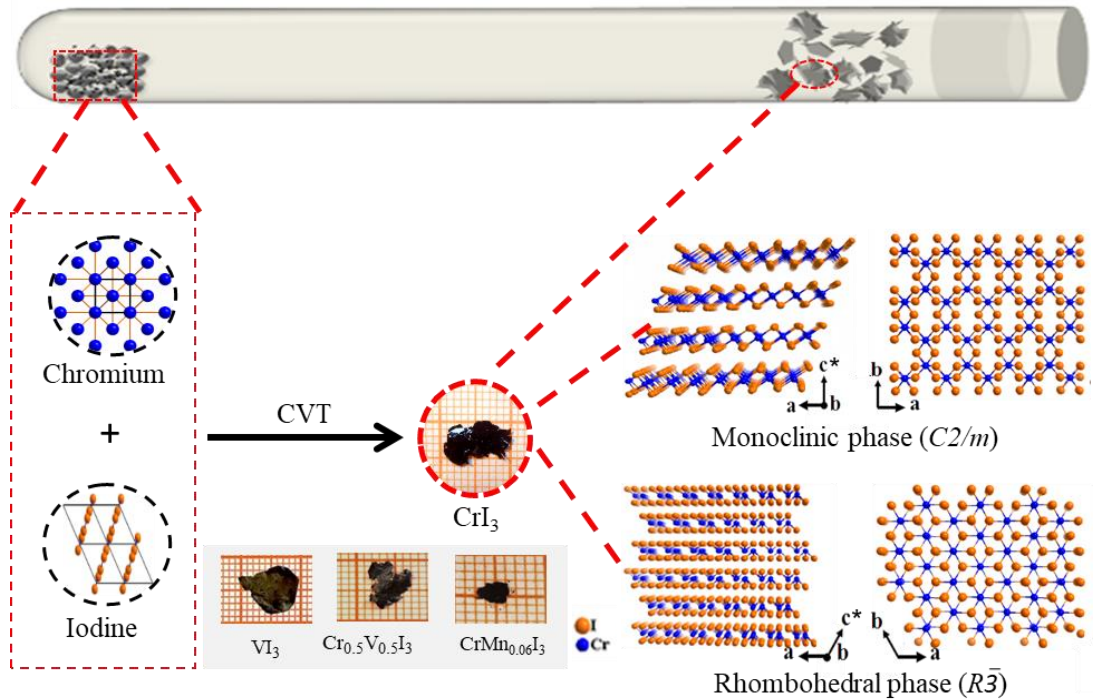

Fig. 1 Schematic illustration of the general synthetic process for CrI$_3$, including VI$_3$, Cr$_{0.5}$V$_{0.5}$I$_3$ and CrMn$_{0.06}$I$_3$, along with their typical crystal structure and general size

**2.2 Measurement and calculation details**

The surface morphology of the samples was observed using a Carl Zeiss optical microscope (Axio Vert. A1). The X-ray powder diffraction (Bruker-AXS D8 Advance) was applied to analyze the phase purity and structure of the samples. The composition and elemental composition were examined using energy dispersive spectroscopy (EDS) in a scanning electron microscope (SEM, FEI Quanta 250F). The magnetic properties are characterized using the Physical Property Measurement System (PPMS, Quantum Design), which measures temperature and magnetic field in the range of 10-300 K and -4-+4 T, respectively.

First principles calculations were performed with the projector augmented wave (PAW) method, as implemented in the Vienna ab initio simulation package (VASP) [27, 28]. The exchange correlation effect was treated with generalized gradient approximation (GGA) function in the form of Perdew-Burke-Ernzerhof (PBE) parametrization. A Hubbard $U$ energy was added to the $d$ electron of V and Cr atom. The static self-consistency calculations were carried out on a $k$ grid of 9×9×9, with the cutoff energy of 600 eV for the plane wave basis set.

3. **Results and discussion**

Fig. 2(a) displays the XRD pattern of $CrI_3$ single crystal collected at 300 K. It is supposed to be the monoclinic structure based on the literature [22], where the peaks can be well indexed to be (*00l*) directions, indicating that the crystal surface is parallel to the *ab* plane. For simplicity, *c\** is applied to represent the direction normal of *ab* plane in the following text. The distinct layered structure of $CrI_3$ is clearly shown by optical microscopy in the two insets of Fig. 2(a). The magnetization loops for both *H//c\** and *H//ab* directions are obtained at 10 K, as shown in Fig. 2(b), where the initial magnetization curve of *H//c\** saturates faster than that of *H//ab*, meaning that the easy magnetization axis is the *c\**-axis. The coercive field along the easy axis is ~200 Oe, implying soft magnetic behavior of $CrI_3$. In contrast to the continuous variation of the magnetic field in *H//ab*, there is a significant magnetization jump at $\mu_0H \approx \pm 2.0$ T for *H//c\**. The calculated jump degree value $\delta = [M(2.2\ T)-M(2.0\ T)]/M(2.0\ T)$ is 2%, similar to that reported in ref. [23, 29]. It is considered as a field-driven AFM to FM transition, where two types of magnetic orders may coexist due to the stacking difference at low temperature, i.e. FM rhombohedral stacking and AFM monoclinic stacking [23, 30]. As the crystals in the present study were grown in a shorter time, the stacking disorder may increase, resulting the obvious jump in the *M-H* curve. This metamagnetic transition can produce high magnetoresistance effects and be applied to spintronic devices, a phenomenon that also appears in other literature and is thought to be common in few-layer $CrI_3$ [30, 31]. However, the presence of a series of jumps around 0.5-1.0 T clearly distinguishes from the experiments of others, which will be explained in detail below. The inset in Fig. 2(b) depicts the temperature dependent magnetization response at $\mu_0H$=0.1 T in both *H//ab* and *H//c\** directions, where a clear anisotropic magnetic response can be observed. The data clearly indicate the FM of $CrI_3$ with a specific $T_c$ close to 57.9 K, which is consistent with the $T_c$ previously reported for thin $CrI_3$ [23]. Fig. 2(c) displays the *c\**-axis magnetization measured at various temperatures (*T*= 10, 30 and 50 K) by sweeping the field between -4 and +4 T. It can be seen that the sub-magnetic jump at $\mu_0H \approx \pm 2.0$T gradually moves to a lower field and eventually disappears at *T*=50 K as the temperature increases.

Fig. 2(d) was obtained by magnifying the *M-H* curve of *H//c\** from Fig. 2(b). The ascending and descending field curves do not overlap at low temperatures, indicating hysteresis behavior. In addition, it is noted that a series of significant discontinuous jumps emerged in the increase of the external magnetic field. Based on the random nature, it is supposed to be the Barkhausen effect, which accompanies the pinning and depinning of the domain walls. The Barkhausen signal in a normal bulk ferromagnetic material is generally small and it was first detected by a voltage pulse on a coil. In the polycrystalline or ferromagnet with high hardness, the dislocation or defect density is relatively high, leading to strong pining of the domain wall, thus Barkhausen activity is not easy to generate and the signal is low [32]. It is assumed the same reason that the Barkhausen effect is absent in previous reported single crystalline $CrI_3$. However, due to the high quality of our sample, the large production of dislocations can be avoided. In addition, the obtained sample is naturally thin, so that the number of the vdW gaps is smaller, facilitating the movement of the domain walls. A simple schematic diagram about the domain wall displacement is shown in Fig. 2(e). The drawn slice is parallel to the *ab* plane, and the magnetization direction is along the easy magnetization axis of $CrI_3$, that is, out of *ab*-plane. The hollow pining sites are sparsely and randomly distributed. In case of applying the magnetic field parallel to the *c\**-axis, the volume of the domains in the same direction gradually increases, and each time the domain wall passes a line of pining sites, it can cause an abrupt jump in the magnetization curve (region from i-ii), until the domain walls finally disappear. The inset in Fig. 2(d) shows similar behavior during sweeping from the negative magnetic field. This behavior of Barkhausen jump, have rarely been observed in previous reports of $CrI_3$ and other similar 2D systems, reflect the high quality and low defect of our samples.

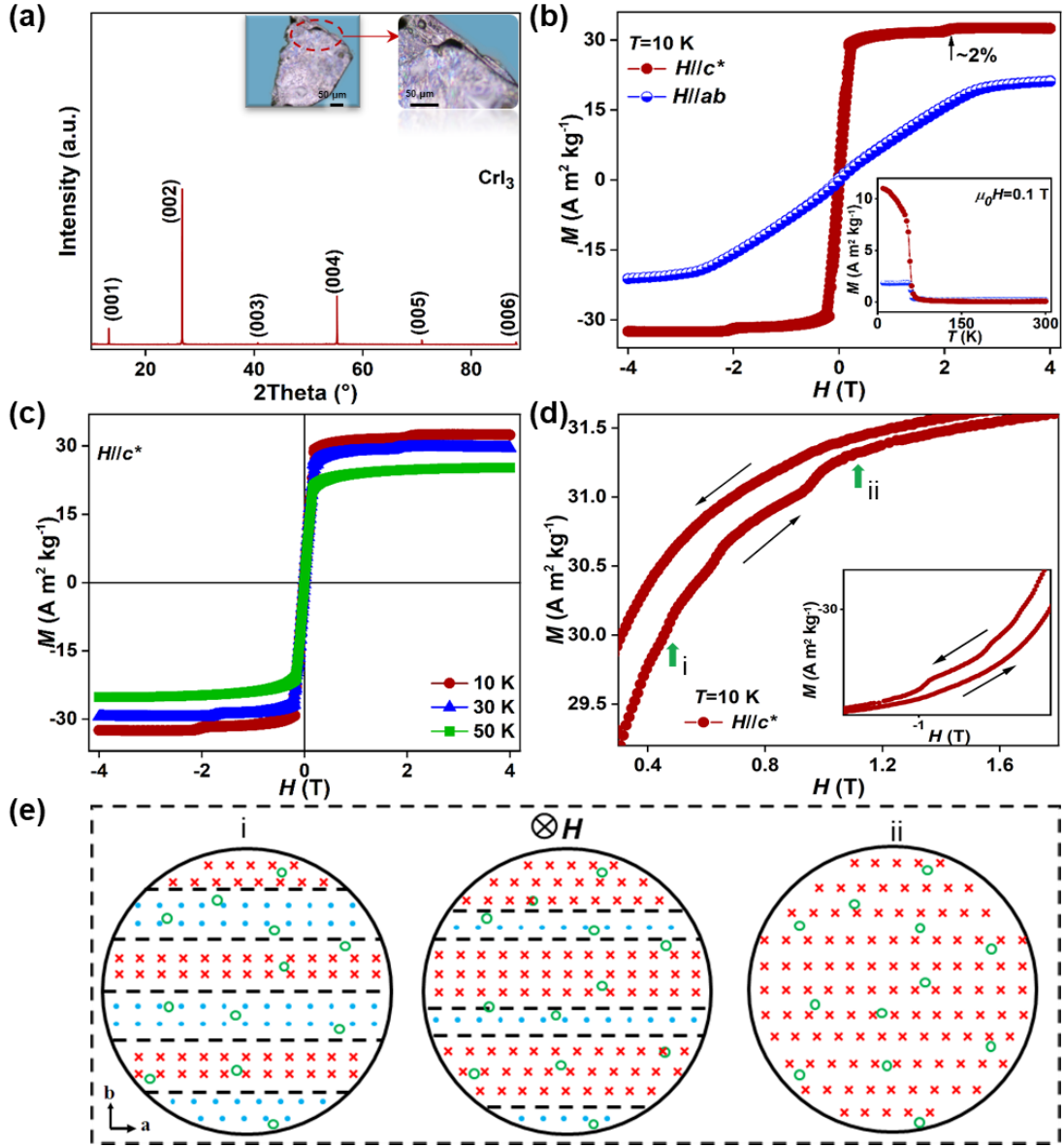

Fig. 2 (a) X-ray power diffraction (XRD) of $CrI_3$. (b) $M(H)$ curves of $CrI_3$ at 10 K for $H//ab$ and $H//c^*$. Inset: $M$-$T$ curves of $CrI_3$ at $\mu_0H$ =0.1 T for $H//ab$ and $H//c^*$. (c) $M$-$H$ curves of $CrI_3$ at 10, 30 and 50 K for $H//c^*$. (d) $M$-$H$ curves of $CrI_3$ at 10 K for $H//c^*$ with a scan field between 0.3 and 1.8 T. Inset: partial enlarged view of $M$-$H$ curves at 10 K. (e) Schematic diagram of $CrI_3$ domain wall displacement.

Then V and Mn element are introduced to the pristine $CrI_3$, aiming to tune the magnetocrystalline anisotropy or Curie temperature of $CrI_3$. The X-ray diffraction patterns of the obtained $Cr_{0.5}V_{0.5}I_3$ and $CrMn_{0.06}I_3$ samples, along with $VI_3$, are shown in Fig. 3(a). All the peak positions highly coincide with those of $CrI_3$, indicating of their similar structures. The interlayer spacing of $VI_3$, $Cr_{0.5}V_{0.5}I_3$ and $CrMn_{0.06}I_3$ also approach to each other, which are around 6.68, 6.47 and 6.43 Å, respectively, within the error of 0.04 Å. The (*002*) peaks of them are enlarged to clearly reveal the evolution, where the position of doped samples slightly shifted to right overall, indicating little shrink of the lattice after doping. Fig. 3(b) shows the temperature dependent magnetization curves of $CrI_3$, $VI_3$, $Cr_{0.5}V_{0.5}I_3$ and $CrMn_{0.06}I_3$ at an applied magnetic field $\mu_0H$=0.1 T, parallel to the *c\**-axis, where their Curie temperatures can be clearly determined. By repeated measurements, the $T_c$ for $CrMn_{0.06}I_3$ is determined to be 57.8±0.1 K, close to that of $CrI_3$, while for $Cr_{0.5}V_{0.5}I_3$, the $T_c$ moves to a lower temperature of 55.0±0.2 K. The $T_c$ of $VI_3$ is lower (~50.2±0.1 K), agreeing with that reported in the literature [33, 34]. Fig. 3(c) compares the *H//c\** magnetization of all compositions at 10 K. It is found that the saturation magnetic moment of $CrI_3$ is about 32 A m$^2$ kg$^{-1}$ (2.5 $\mu_B$), which is lower than its theoretical value (~3.1 $\mu_B$) [35]. The possible reason is the degradation of thin layer $CrI_3$ in the air, causing the increase of its real mass, thus resulting in decrease of the magnetic moment. It is interesting to note that despite the low doping efficiency, $CrMn_{0.06}I_3$ exhibit a much larger saturation magnetic moment about 44 A m$^2$ kg$^{-1}$ (3.4 $\mu_B$) comparing to $CrI_3$. Considering the largest magnetic moment of Mn ion, that is 4 $\mu_B$, the increment in $CrMn_{0.06}I_3$ can only reach 0.24 $\mu_B$. Therefore, it is assumed that the Mn doping may also affect the structure of $CrI_3$ and enhance the average magnetic moment of Cr ion. Further, the isothermal magnetization curves of $CrMn_{0.06}I_3$ at different temperatures in Fig. 3(d) exhibit similar reversible jump at ~±2.0 T. The Barkhausen signal can also be observed at 10 K, but with weaker amplitude.

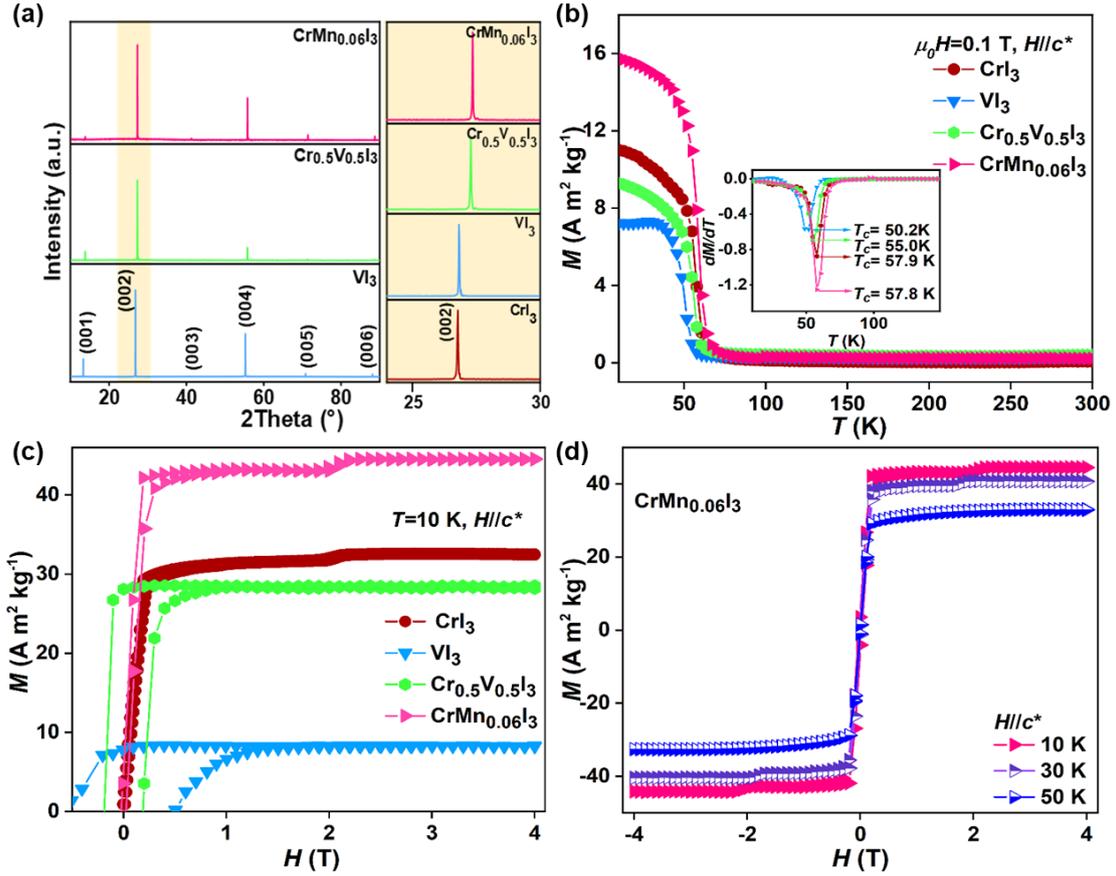

Fig.3 (a) XRD of $VI_3$, $Cr_{0.5}V_{0.5}I_3$ and $CrMn_{0.06}I_3$ single crystals and (*002*) enlarged diffraction peaks for all peaks. (b) *M-T* curves along $H//c*$ at $\mu_0H$ =0.1 T for all samples. Inset: *dM/dT* vs *T* for all samples. (c) *M-H* curves for them at 10 K for all samples. (d) Isothermal magnetization curves of $CrMn_{0.06}I_3$ for *T*=10, 30 and 50 K measured in the out-of-plane field.

For V doped sample, the saturation magnetization lowers, due to the smaller moment of V ions, as seen in the following part of calculation. However, the remnant magnetization grows, as with increasing V doping, the coercive field grows substantially from ~200 to ~4600 Oe, as shown in Fig. 4(a). Fig. 4(b) shows the in- and out-of-plane *M-H* curves of $Cr_{0.5}V_{0.5}I_3$ at 10 K. The anisotropic field is expected to be greater than 4 T and much higher than $CrI_3$. Therefore, the increasing coercivity can be related with growing magnetocrystalline anisotropy in V doped $CrI_3$. For the middle composition of $Cr_{0.5}V_{0.5}I_3$, the $H_c$ is about 1880 Oe, and it gradually decreases with increasing temperature, as seen in the inset of Fig. 4(b). Moreover, after V doping, the AFM-FM sub-magnetic transition is disappeared. It may be caused by the stable FM exchange interaction in $VI_3$, whether in few layers or bulk [36].The Barkhausen jumps also vanish, accompanying with the enlargement of $H_c$. The hard magnetic properties of $Cr_{0.5}V_{0.5}I_3$, while similar $T_c$ and saturation magnetization, make it a better choice comparing to CrI$_3$ when constructing spintronic devices.

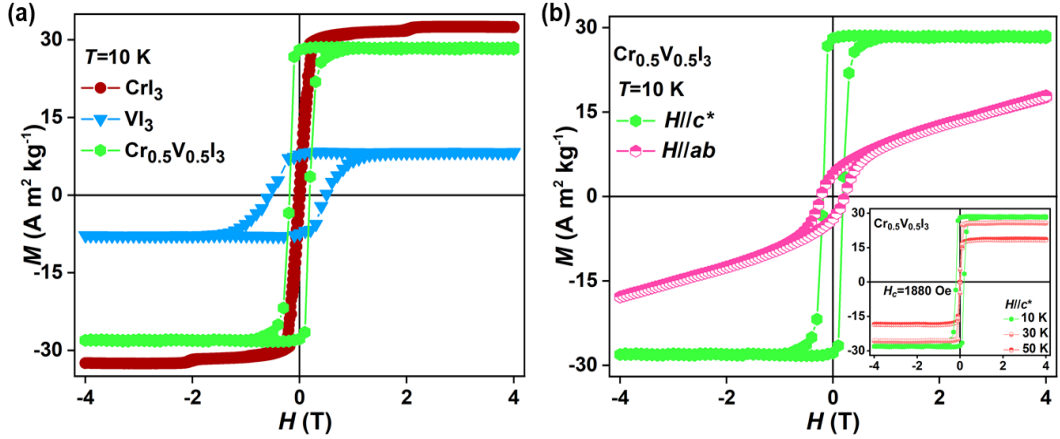

Fig. 4 (a) *M-H* curves of CrI$_3$, VI$_3$ and Cr$_{0.5}$V$_{0.5}$I$_3$ at *T* =10 K for *H*//*c**. (b) *M-H* curves of Cr$_{0.5}$V$_{0.5}$I$_3$ for *H*//*c** and *H*//ab. Inset: *M-H* curves of Cr$_{0.5}$V$_{0.5}$I$_3$ at 10, 30 and 50 K for *H*//*c**.

Finally, the electronic structures of the V doped sample are also investigated by first principles calculations. It is found that the band structure of VI$_3$ presents an incorrect metallic state without considering the electron-electron correlation Hubbard U energy, suggesting that VI$_3$ is a FM Mott insulator[37]. Based on the literature[38], a Hubbard U energy of 3.68 eV was applied to the d-electron of V atom. For CrI$_3$, we find the inclusion of U barely altered the band structure. For consistency, a common U value of 2.5 eV was still added to Cr. As shown in Fig. 5, they all exhibit insulating state, with a rather flat d-band lying across the Fermi level. It is noted that the indirect band gap in CrI$_3$ turns to direct ones in Cr$_{0.5}$V$_{0.5}$I$_3$ and VI$_3$ samples, accompanying with a decreasing bandgap magnitude. Obvious stronger spin-splitting is observed for CrI$_3$ across the Fermi level, corresponding to a lager magnetic moment comparing with VI$_3$.

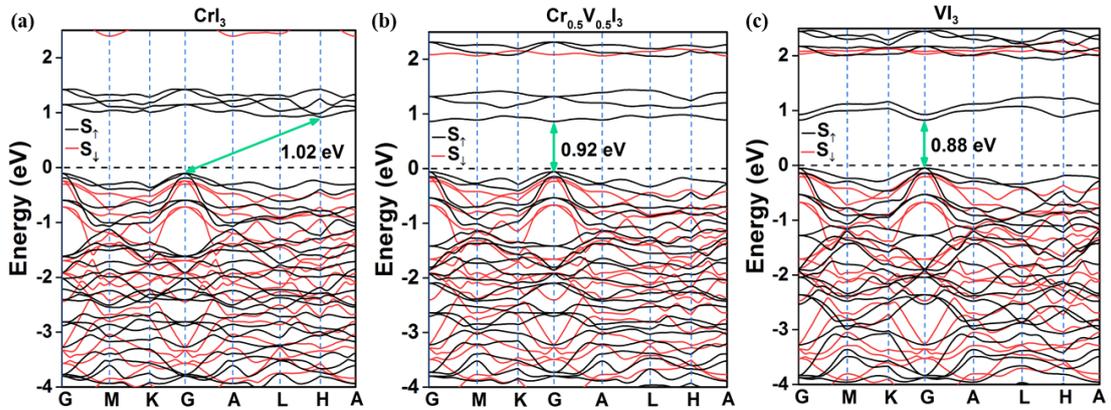

Fig. 5 Spin resolved band structure of CrI$_3$ (a), Cr$_{0.5}$V$_{0.5}$I$_3$ (b) and VI$_3$ (c). Their bandgap values are also indicated in the figure.

## 4. Conclusion

In conclusion, we have successfully synthesized large-size and high-quality $CrI_3$, $VI_3$, $Cr_{0.5}V_{0.5}I_3$ and $CrMn_{0.06}I_3$ single crystals by optimization of the preparation process and investigated their structural and magnetic properties. Remarkable irreversible Barkhausen jump was observed in the *M-H* curves of thin $CrI_3$ crystal. The low dislocation density could possibly account for it. The large Barkhausen signal in layered $CrI_3$ can offer a good platform to investigate the mechanism of this effect. Besides, the specific domain motion could be interesting due to the layered structure and is worth to be investigated in more detail by future in-situ domain observation or micro-magnetic simulations. By small doping of Mn into $CrI_3$, the saturation magnetization can substantially enhance, while the magnetization process is almost maintained. For V doping, both the magnetocrystalline anisotropy and coercivity significantly grows, making $Cr_{0.5}V_{0.5}I_3$ a good hard magnet, while keeping similar $T_c$ and good environmental-stability as $CrI_3$. By analyzing their electronic structures with first principles calculations, it is found that the indirect band gap in $CrI_3$ turns to direct ones in $Cr_{0.5}V_{0.5}I_3$ and $VI_3$, which also exhibit smaller gap magnitude. Moreover, the V contained sample exhibit large electron correlation effect, indicating great modulation possibility of their transport properties.

## Acknowledgements

This work is supported by National Natural Science Foundation of China (Grant No. 11604148), and the Open Fund of Large Facilities in Nanjing University of Science and Technology.